\documentclass[showpacs,aps,twocolumn]{revtex4}
\usepackage{epsfig}
\usepackage{graphicx}
\usepackage{amsmath,amssymb,amsfonts}
\usepackage{array}
\usepackage{url}
\usepackage{hyperref}
\usepackage{multirow}
\usepackage{float}
\usepackage{lineno}
\usepackage{xspace}
\usepackage[usenames,dvipsnames]{color}

\newcommand{\bef}{\begin{figure}}
\newcommand{\eef}{\end{figure}}
\newcommand{\bc}{\begin{center}}
\newcommand{\ec}{\end{center}}

\newcommand{\be}{\begin{equation}}
\newcommand{\ee}{\end{equation}}
\newcommand{\bea}{\begin{eqnarray}}
\newcommand{\eea}{\end{eqnarray}}

\def\ba{\begin{eqnarray}}
\def\ea{\end{eqnarray}}
\begin{document}

\title{Characterizing Proton-Proton Collisions at the Large Hadron Collider with Thermal Properties}
\author{Dushmanta Sahu}
\author{Raghunath Sahoo\footnote{Corresponding Author Email: Raghunath.Sahoo@cern.ch}\footnote{Presently CERN Scientific Associate at CERN, Geneva, Switzerland}}
\affiliation{Department of Physics, Indian Institute of Technology Indore, Simrol, Indore 453552, India}

\begin{abstract}
High-multiplicity proton-proton (pp) collisions at the Large 
 Hadron Collider (LHC) energies have created a new domain of research to 
look for a possible formation of quark--gluon plasma in these events. 
In~this paper, we estimate various thermal properties of the matter formed 
in pp collisions at the LHC energies, such as mean free path, isobaric 
expansivity, thermal pressure, and heat capacity using a thermodynamically 
consistent Tsallis distribution function.
 %, which happens to describe the identified particle 
 %transverse momentum spectra very well. 
 Particle species-dependent mean free path and isobaric expansivity are 
studied as functions of final state charged particle multiplicity for 
pp collisions at the center-of-mass energy $\sqrt{s}$ = 7 TeV. 
The effects of degree of non-extensivity, baryochemical potential, and 
temperature on these thermal properties are studied. The~findings are 
compared with the theoretical expectations.
\end{abstract}
\date{\today}
\maketitle{}

\section{Introduction}
\label{intro}
One of the many astounding revelations of 20th century physics is the 
relatively new state of matter called quark--gluon plasma (QGP), expected 
to be formed in ultra-relativistic heavy-ion collisions. The~temperature 
in such collisions reaches up to a few hundred MeV. For a better grasp of 
this temperature, consider the temperature at the core of the Sun: 
the~temperature at the core of the Sun is around $10^{5}$ times less than 
that produced in heavy-ion collisions at Relativistic Heavy Ion Collider 
(RHIC) and the Large Hadron Collider (LHC). At~such high temperatures, the~normal baryonic matter with hadrons as their degrees of freedom goes through a cross-over phase transition to produce QGP, which has quarks and gluons as its degrees of freedom. Earlier,  the~general understanding was that only high-energy heavy-ion collisions can produce QGP while proton-proton 
(pp) collisions were taken as baseline measurements to understand medium formation 
in nuclear collisions. However, recent advances in high-energy pp collisions have shown us that there is a possibility to form QGP droplets, hopefully in high-multiplicity pp collisions in view of the observed heavy-ion-like signatures in such collisions~\cite{nature,Velicanu:2011zz}. Perturbative QCD (pQCD)-based Monte Carlo event generators like PYTHIA8 model in their advanced tunes such as multipartonic interactions, color reconnection, and rope hadronization have been successful in explaining some of the features in small collision systems~\cite{Pythia8}. These include color reconnections bringing out flow-like features in small systems~\cite{Ortiz:2013yxa}, which is in principle a macroscopic hydrodynamic feature with a possible microstate dynamics responsible for thermalization, and rope hadronization explaining the enhancement of 
multi-strange particles with respect to pions~\cite{Nayak:2018xip}. Although~there is no built-in thermalization in PYTHIA8, it is successful 
in mimicking the macroscopic flow features in pp collisions. 
In contrast to PYTHIA8, the energy conserving quantum mechanical multiple 
scattering approach, based on partons (parton ladders), off-shell 
remnants, and Splitting of parton ladders EPOS Monte Carlo model also describes the heavy-ion-like features in small collision systems~\cite{Werner:2013tya}. It includes event-by-event 3+1-dimensional hydrodynamic evolution of the system produced in high-energy collisions~\cite{Werner:2013tya,Werner:2010aa,Drescher:2000ha}. The~description of hard sectors for PYTHIA8 and EPOS generator is similar; however, the~key difference in the EPOS model is that it includes a hydrodynamical description of the underlying mechanisms for generating collectivity instead of the color reconnection mechanism in PYTHIA8. Therefore, unlike PYTHIA8, EPOS includes the thermalization of a core (bulk) part of the~system.

In view of these, there is a need to obtain a deeper understanding of high-multiplicity pp events to obtain a better understanding of the produced systems, possible associated thermodynamics, particle production dynamics, and freeze-out. There have been attempts to study these events using event topology, which separates jetty
(pQCD-based hard scattering events) from isotropic (dominated by soft 
processes) events~\cite{Acharya:2019mzb,Khatun:2019dml,Cuautle:2015kra,Rath:2019izg,Tripathy:2019blo,Khuntia:2018qox}. Using the temperature, $T$, and Tsallis non-extensive parameter, $q$, extracted from the transverse momentum  $p_\perp$-spectra, various thermodynamical quantities are estimated in order to characterize the system created in pp~\cite{Deb:2019yjo} and heavy-ion collisions~\cite{Azmi:2019irb}. In line with these explorations from
various fronts, in the present study, 
we make an attempt to understand some of the thermodynamic features of pp collisions using 
experimentally motivated Tsallis non-extensive distribution, which describes the particle spectra \cite{Bhattacharyya:2017hdc,Khuntia:2017ite}. In our study, we consider a cosmological 
expansion scenario, where the fireball produced in high-energy hadronic and nuclear collisions cools down during its spacetime
expansion following a temperature profile with time \cite{Tripathy:2019blo}. 
The final state particles decouple from the system following a differential freeze-out
scenario, with heavier particles coming out of the system early in time, which corresponds to higher decoupling temperatures. These 
decoupling points also depend on the interaction of cross section of individual particles with the produced~system.  

To understand the behavior of the matter formed in high-energy collisions at ultra-relativistic energies, we need to obtain an overall idea about its thermodynamical properties. The mean free path of a system gives the average distance that a particle has to travel before it collides with another particle in the system. It can give us an idea about the state of the system under consideration. One of the important thermodynamic properties of a system to study is isobaric expansivity, $\alpha$. The~tendency of matter to change its shape, area, or volume is known as thermal expansion of that matter. The~relative expansion per change in temperature at constant pressure is called the matter's coefficient of thermal expansion or isobaric expansivity~\cite{KHuang}. The~study of isobaric expansivity of the matter formed in high-energy collisions tells us about the key features in such systems. Thermal pressure, $\frac{\partial P}{\partial T}$, is nothing but the ratio of isobaric expansivity and isothermal compressibility. It tells us how much the pressure of the system increases with an increase in temperature at constant volume. The heat capacity, $C_{\rm V}$, of the system
  is the amount of heat energy that is required to raise the temperature of the system by one unit at constant volume. It tells us how the entropy, $S$, of the system changes with the change in temperature: $\Delta S = \int_{}^{}\frac{C_{\rm V}}{T}dT$. Entropy being directly related
  to the experimentally measured midrapidity charged particle density in pseudorapidity, $dN_{\rm ch}/d\eta$, is an important observable when studying the QCD phase transition \cite{Kharzeev:2017qzs}. This makes the study of $C_{\rm V}$ more~interesting. 
  
In high-energy collisions, a large number of final state particles are produced, which encourage us  to take a statistical approach to describe such systems. The~transverse momentum 
of the produced particles are expected to follow a thermalized Boltzmann--Gibbs (BG) distribution function. However, there have been observations of a finite degree of deviation from the equilibrium statistical description of identified particle $p_\perp$-spectra in experiments at RHIC~\cite{Abelev:2006cs,Adare:2011vy} and LHC~\cite{Aamodt:2011zj,Abelev:2012cn,Abelev:2012jp,Chatrchyan:2012qb}. This deviations are a result of a higher contribution of pQCD processes and can be better described by a combination of Boltzmann-type distribution functions along with a pQCD inspired power-law function \cite{Bhattacharyya:2015hya}. Tsallis non-extensive statistical distribution has been found to be very accurate in explaining the $p_\perp$-spectra of hadronic collisions and is used to obtain particle multiplicities in experimental measurements \cite{Abelev:2006cs,Adare:2011vy,Aamodt:2011zj,Abelev:2012cn}. 
Whereas Tsallis distribution function fits the $p_{\perp}$-spectra very 
well in the 
midrapidity region, there are other types of distribution functions in the literature that can fit the $p_\perp$-spectra away from zero rapidity very well~\cite{Pirner:2011ab}.
As a Boltzmann-type distribution, which is isotropic in transverse and longitudinal momenta, cannot explain the whole $p_\perp$-spectra, in the present study, 
we use a thermodynamically consistent Tsallis distribution function to describe the $p_\perp$-spectra in pp collisions~\cite{Cleymans:2011in}.
The~parameter $q$ in Tsallis distribution function denotes the degree of 
deviation from the equilibrium, and $q$ = 1 corresponds to the equilibrium 
condition (BG scenario). At~higher charged particle multiplicities in 
heavy-ion collisions, the~value of $q$ tends towards 1, which signifies 
that the system has obtained thermodynamic equilibrium. From~the Tsallis distribution fit to the $p_\perp$-spectra of pp collisions at the centre-of-mass energy $\sqrt{s}$ = 7 TeV, we extracted thermodynamical parameters such as temperature, $T$, and the non-extensive parameter, $q$, which were then used to find the particle species-dependent mean free path and isobaric expansivity of the hadronic system formed in pp collisions.

The paper is organized as follows. Section~\ref{formulation} gives a brief account of the formulations  used in the paper.
In~Section~\ref{res},  we discuss our findings and the results, and~in Section~\ref{sum}, we summarize our investigation with
important~findings. 

\section{Formulation}
\label{formulation}
As the identified particle $p_\perp$-spectra for pp collisions are well-described by a Tsallis non-extensive distribution function, we estimated various thermal properties of the produced fireball using a thermodynamically consistent Tsallis distribution function, which is given as follows~\cite{Cleymans:2011in,Cleymans:2012ya,Deppman,Deppman1}:
\begin{equation}
\label{eq1}
f^q (E, q, T, \mu) = \frac{1}{\bigg [1 + (q-1)\frac{E - \mu}{T} \bigg]^\frac{q}{q-1}},
\end{equation}
where $T$ and $\mu (= \mu_{\rm B}+ \mu_{\rm Q} + \mu_{\rm S})$ are the temperature and the chemical potential, respectively, with $\mu_{\rm B}$ as the baryochemical potential, $\mu_{\rm Q}$ as the chemical potential related to the electric charge and $\mu_{\rm S}$ as the chemical potential related to the strangeness quantum number. Here, $E=\sqrt{p^{2}+m^{2}}$ is the energy, with $p$ being the momentum and $m$ being the mass of the particle under study. The~thermodynamical quantities in non-extensive statistics are given by,
\begin{equation}
\label{eq2}
n = g \int \frac{d^{3}p}{(2\pi)^3}\bigg[1 + (q-1)\frac{E- \mu}{T} \bigg]^\frac{-q}{q-1}\, ,
\end{equation}
\begin{equation}
\label{eq3}
\epsilon = g \int \frac{d^{3}p}{(2\pi)^3} E \bigg[1 + (q-1)\frac{E- \mu}{T} \bigg]^\frac{-q}{q-1}\, ,
\end{equation}
\begin{equation}
\label{eq4}
P = g \int \frac{d^{3}p}{(2\pi)^3} \frac{p^{2}}{3E} \bigg[1 + (q-1)\frac{E- \mu}{T} \bigg]^\frac{-q}{q-1}\, .
\end{equation}
where $n$,
 $\epsilon$, $P$, and $g$ are the number density, energy density, pressure, and degeneracy, respectively.
This formulation gives a thermodynamically consistent description, meaning the~basic thermodynamic
relationships are satisfied:
\begin{eqnarray}
\label{thermo}
d\epsilon = T~ds + \mu~ dn,\\
dP=s~dT +n~d\mu,\\
\epsilon + P = Ts +\mu n.
\end{eqnarray}

Furthermore,
 the~variable $T$, used in the above expressions, obeys the thermodynamic~relation:
\begin{equation}
\label{temp}
T = \frac{\partial U}{\partial S}\bigg |_{N,V},
\end{equation}

Hence, it could be termed as the temperature even though the system follows Tsallis non-extensive statistics. Here, $U$ is the internal energy 
of the system. It is noteworthy to mention here that there are various forms of Tsallis distribution functions available in the literature~\cite{Biro:2020kve,Deppman:2017fkq}. However, the~form used in this work is thermodynamically consistent and, hence, gives a scope to deal with various thermodynamic variables to characterize the
created~fireball. 

Going ahead with the above prescription, in~view of thermodynamic consistency and Tsallis non-extensive statistics, the~expressions for the discussed thermodynamic
observables could be written as follows. The~general expression of mean free path is given by
\begin{equation}
\lambda = \frac{1}{n\sigma}.
\end{equation}
Here $\sigma$ is the scattering cross section. Using Equation~(\ref{eq2}), the~expression for mean free path for the hadron gas becomes
\begin{equation}
\label{eq5}
\lambda = \frac{1}{\sum_{a} \sigma g_{a}\int_{}^{}\frac{d^3p}{(2\pi)^3} \lbrack 1+(q-1)\frac{E_{a}-\mu}{T}\rbrack^{\frac{-q}{q-1}} },
\end{equation}
where $E_{a}$ is the energy of the $a$th particle and $\sigma$ can be taken as 11.3 mb considering a 
hard core hadron radius, $r_h$ = 0.3 fm, and using $\sigma = 4\pi r_h^2$~\cite{Kadam:2015xsa,Bugaev}.  The~
choice of an average value of hard core hadron radius, 
 $r_h$ = 0.3 fm, seems to be a good choice in the estimation
of viscosity of the system~\cite{Kadam:2015xsa} while giving proper estimation of hadron yields in a statistical hadron gas 
model~\cite{Bugaev}. A~change in $r_h$ in the calculation of hadron scattering cross section may induce
a change in the absolute values of $\lambda$. However,  we have checked it explicitly that, this will have no impact on the trend of $\lambda$ of the
particles when it is studied as a function 
of temperature, final state multiplicity, $q$-parameter, or the baryochemical potential of a~system. 

From thermodynamics, the~expression of isobaric expansivity is given as~\cite{KHuang}:
\begin{equation}
\label{eq6}
\alpha = \frac{1}{V}\frac{\partial V}{\partial T}\bigg \vert_{P}.
\end{equation}

As the exact freeze-out volume for identified charged particles is not very well measurable, Equation~(\ref{eq6}) can be rewritten in terms of particle number density, $n$ as follows:

\begin{equation}
\alpha = -\frac{1}{n}\frac{\partial n}{\partial T}\bigg \vert_{P}.
\end{equation}

Using Equation~(\ref{eq2}), one obtains

\begin{equation}
\label{eq7}
\alpha = -\frac{\sum_{a} \int_{}^{} q\lbrack 1+(q-1)\frac{E_{a}-\mu}{T}\rbrack^{\frac{1-2q}{q-1}}(\frac{E_{a}-\mu}{T^2})d^3p}{\sum_{a} \int_{}^{} \lbrack 1+(q-1)\frac{E_{a}-\mu}{T}\rbrack^{\frac{-q}{q-1}} d^3p}.
\end{equation}

The thermal pressure is given by (see Equation~(\ref{eq4})) 
\begin{equation}
\label{eq12}
\bigg(\frac{\partial P}{\partial T} \bigg)_{\rm V} = \sum_{a} g_{a}\int_{}^{}\frac{d^3p}{(2\pi)^3} \frac{qp^2}{3E_{a}}\bigg \lbrack 1+(q-1)\frac{E_{a}-\mu}{T}\bigg \rbrack^{\frac{1-2q}{q-1}}
\bigg(\frac{E_{a}-\mu}{T^2}\bigg).
\end{equation}

The heat capacity or the specific heat at constant volume is defined as
\begin{equation}
C_{\rm V} = \frac{\partial \epsilon}{\partial T}, 
\end{equation}
and, using Equation~(\ref{eq3}), this is given by

\begin{equation}
\label{eq13}
C_{\rm V} = \sum_{a}g_{a}\int_{}^{}\frac{d^3p}{(2\pi)^3} qE_{a}\bigg \lbrack 1+(q-1)\frac{E_{a}-\mu}{T}\bigg \rbrack^{\frac{1-2q}{q-1}}\bigg(\frac{E_{a}-\mu}{T^2}\bigg).
\end{equation}

In all of the above equations, we have taken $\mu = \mu_B$, assuming exact charge and strangeness conservation in the system.
In the present study, we took a particle mass cutoff of 1.5 GeV to include all the hadrons in the Review of Particle Physics \cite{PDG} in all of our theoretical estimations. In~LHC pp collisions, the baryochemical potential is assumed to be zero because~of the baryon-antibaryon symmetry. Therefore, in our calculations, we took $\mu_B$ = 0 except~when we study the effect of baryochemical potential on various thermal observables. There are recent developments ~\cite{Cleymans:2020ojr}, where $\mu_B$ is extracted as a fit parameter from the Tsallis fit to the transverse momentum spectra, while~this is beyond the scope of the
present study.

%%%%%%%%%%%%%%%%%%%%%%%%%%%%%%%%%%%%%%%%%%%%%%%%%%%%%%%%%%%%%%%%%%%%%%%%%%%%%%

\section{Results and~Discussion}
\label{res}
\begin{figure}
%\begin{center}
\includegraphics[scale = 0.45]{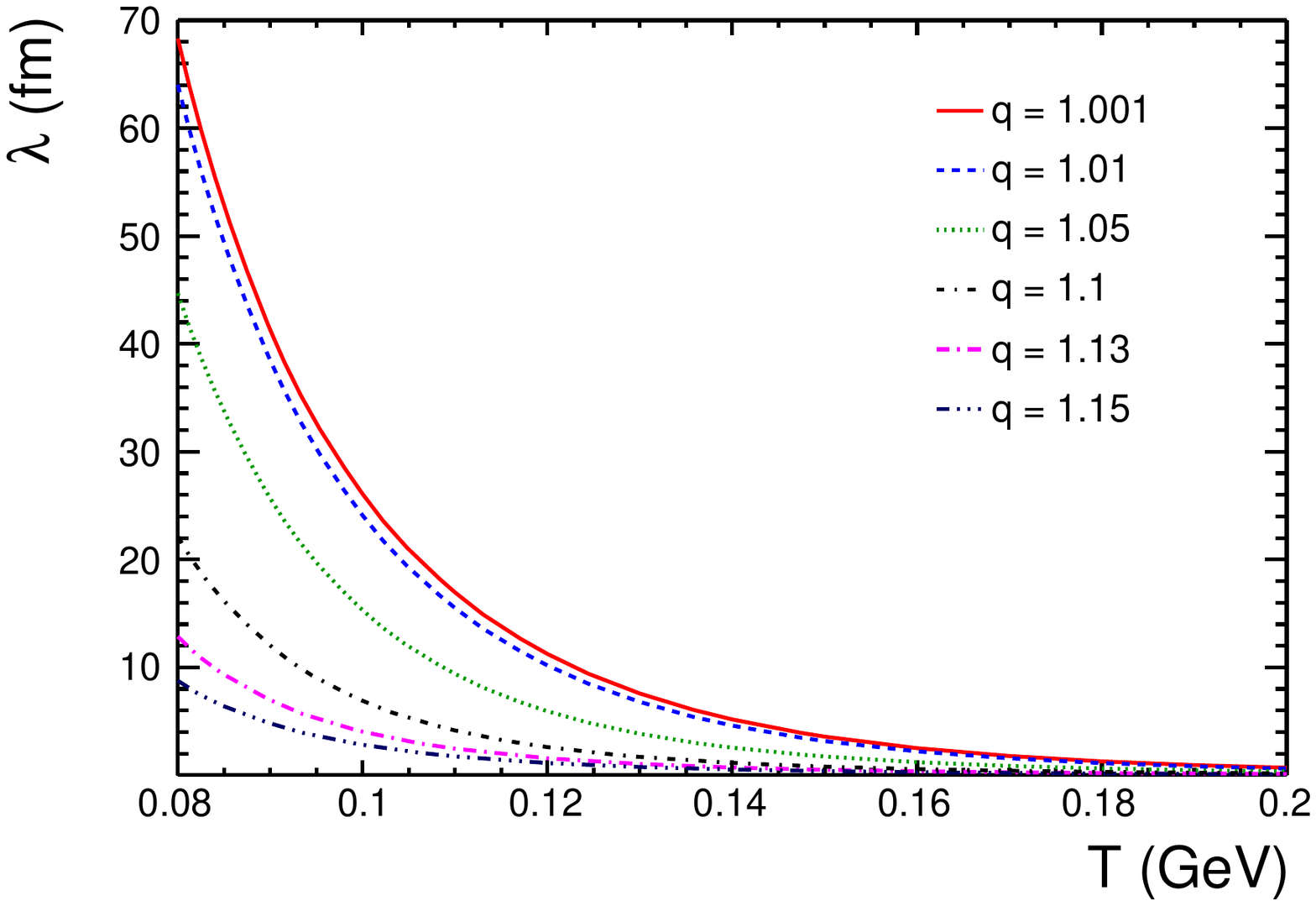}
\caption{Mean free path, $\lambda$, as a function of temperature, $T$, for 
a hadron gas for different values of non-extensive parameter, $q$, for 
baryochemical potential, $\mu_{\rm B}$ = 0.}
\label{fig1}
%\end{center}
\end{figure}

To calculate the mean free path, we used Equation~(\ref{eq5}). In~Figure~\ref{fig1}, we plot the mean free path, $\lambda$, of a hadron gas system as a function of temperature for different values of $q$-parameter. One observes that, for low temperature, the~mean free path of the system is very high and suddenly decreases, becoming minimum at the high temperature region, which usually
corresponds to freeze-out temperatures in high energy hadronic and nuclear collisions~\cite{Kraus:2007hf}. One also notices
that the mean free path is highly sensitive to
the $q$ parameter in the low temperature regime, whereas the dependency decreases at higher temperatures becoming almost negligible. These trends are expected because, at~low temperature, the number density in the system is less, which results in a high $\lambda$. As~the temperature increases, the~number density also increases drastically between a short temperature range before becoming almost constant at very high temperatures. In~the low temperature regime, at~a given temperature, the~mean free path decreases when the system slowly moves away from equilibrium. For~an expanding fireball, initially, the system volume is low and, hence, for a given number of particles, the mean free path is smaller.
The constituents of the system go through mutual collisions as a part of the process of equilibration, while the system is still 
expanding, resulting in a higher mean free path. Below~the expected kinetic freeze-out temperature, the~fireball volume becomes much higher and the mean free path becomes higher than the system volume.

This means that, at~this point and beyond, the~momentum spectra are frozen because the constituent particles hardly collide with each~other.

\begin{figure}
%\begin{center}
\includegraphics[scale = 0.45]{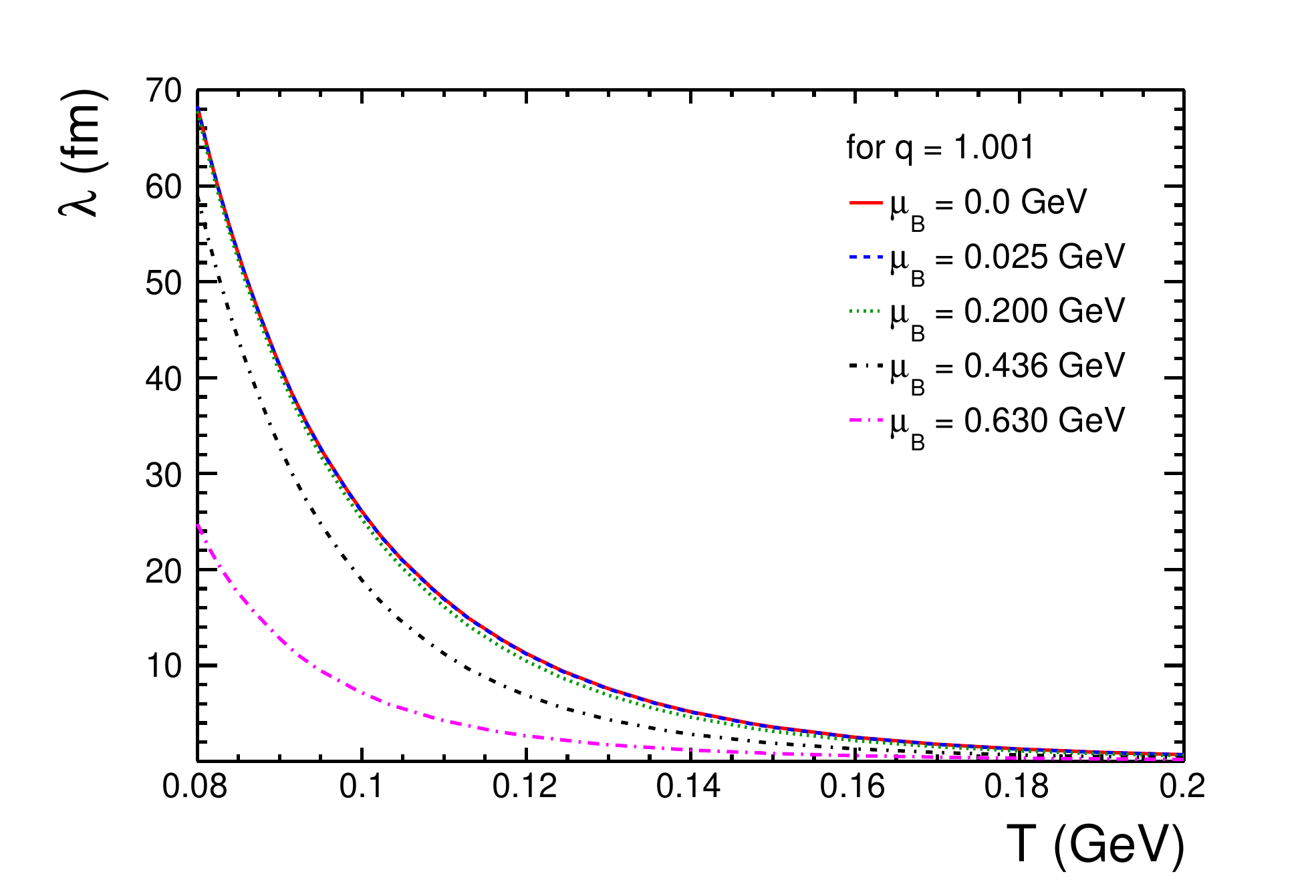}
\caption{Mean free path, $\lambda$, as a function of temperature, $T$, for 
a hadron gas for different values of baryochemical potential, $\mu_{\rm 
B}$.} 
\label{fig2}
%\end{center}
\end{figure}

Figure~\ref{fig2} shows the mean free path of a hadron gas as a function 
of temperature for different values of baryochemical potentials. For~this 
study, we took five different baryochemical potentials; $\mu_{\rm B}$ = 0, 0.025, 0.200, 0.436, and 0.600 GeV, which correspond to, respectively, LHC energies, 
RHIC at $\sqrt{s_{NN}}$ = 200 GeV and 19.6 GeV, RHIC/FAIR at $\sqrt{s_{NN}}$ = 7.7 GeV, and NICA at $\sqrt{s_{NN}}$ = 3 GeV~\cite{Tawfik:2016sqd,BraunMunzinger:2001ip,Cleymans:2005xv,Khuntia:2018non}.
Here, we took the non-extensive parameter, $q$ = 1.001, which corresponds to a near equilibrium state. One observes that, when $\mu_{\rm B}$ is zero, the~trend of the mean free path of the system mimics the trend which we get from Figure~\ref{fig1}. However, as~we go on increasing the value of $\mu_{\rm B}$, which happens when one moves down in collision energy, the~mean free path of the system at a given temperature becomes lower. The~dependence on baryochemical potential on the mean free path is 
clearly visible in the low temperature regime. There is  universality in the decreasing nature of the mean free path as a function 
of temperature for various $\mu_{\rm B}$ values. This is due to the fact that, with an increase in baryon chemical potential, the~number density of the system will increase, resulting in a lower mean free path.
Similarly, one expects to produce more particles at higher system temperatures, making the number density in the 
system higher. This makes the mean free path decrease with temperature irrespective of the baryochemical potential of the
system. This is clearly evident from Figure~\ref{fig2}. 

\begin{figure}
%\begin{center}
\includegraphics[scale = 0.45]{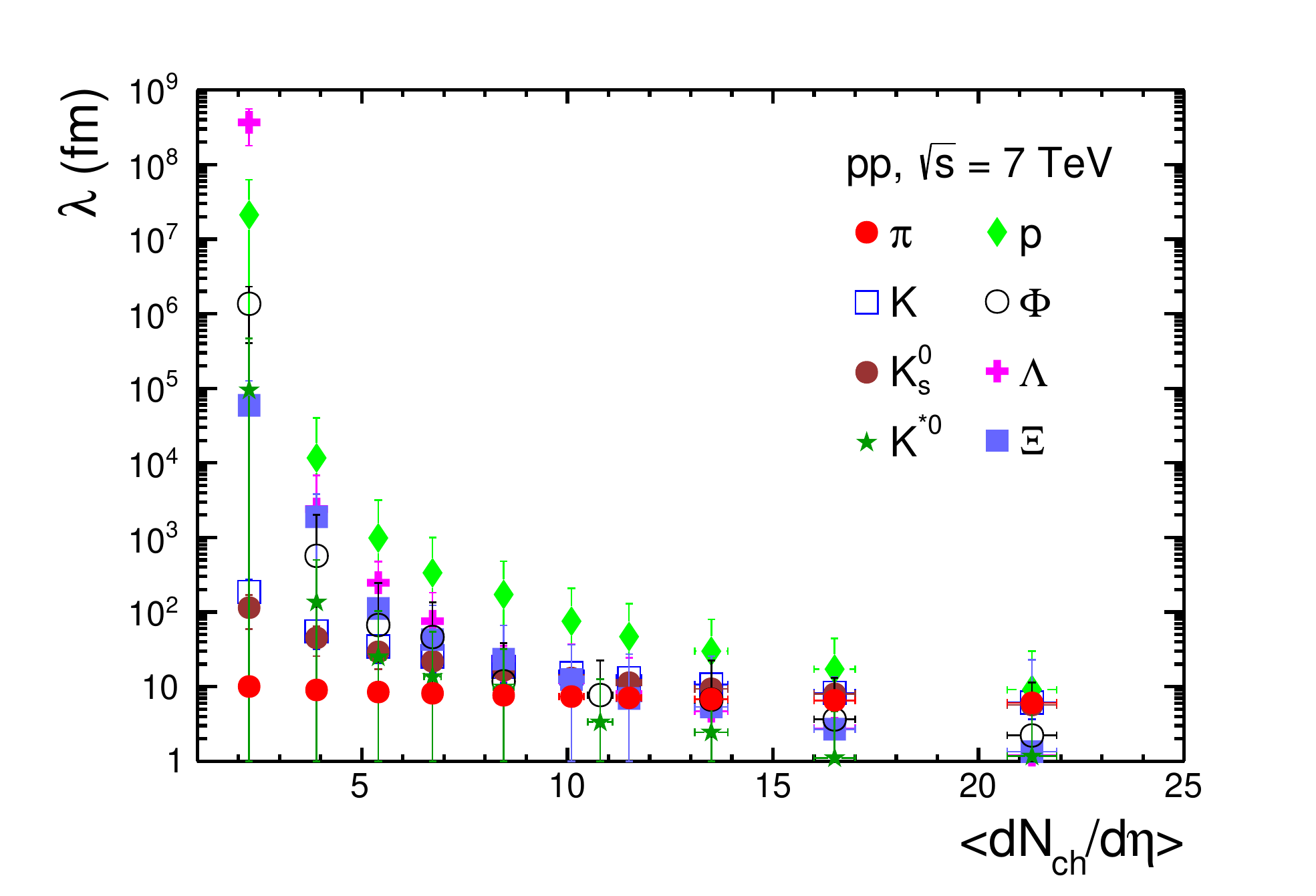}
\caption{Mean free path, $\lambda$, as a function of final state charged 
particle mean multiplicity density in pseudorapidity, $\langle dN_{\rm 
ch}/d\eta\rangle$ (which is a centrality classifier), for different identified particles in pp collisions at $\sqrt{s}$ = 7 TeV. The values
of the mean free path are estimated by using Equation (\ref{eq5}) by 
taking 
the $(T,q)$ values from the Tsallis distribution fitting to the experimental transverse momentum spectra of identified particles.}
\label{fig2-1}
%\end{center}
\end{figure}

Beyond these studies on the dependencies of mean free path on the $\mu_{\rm B}$, $T$, and $q$ parameters of the system for a hadron
gas, one must be curious to look into the behaviour of mean free path as a function of final state event multiplicity for pp collisions 
at LHC energies for different identified particles. In~order to do that, we took the multiplicity-dependent identified particle
$p_\perp$-spectra for pp collisions at $\sqrt{s}$ = 7 TeV, and using the discussed Tsallis non-extensive statistics, we
obtained the $T$ and $q$ parameters~\cite{Khuntia:2018znt}. Taking these as inputs, we estimated $\lambda$ as a function of final state event multiplicity,
which is shown in Figure~\ref{fig2-1}. The~definition of event class through final state multiplicity was taken from the ALICE experiment~\cite{nature}. The~hardening of $p_\perp$-spectra with event multiplicity for pp collisions at $\sqrt{s}$ = 7 TeV has been
seen recently~\cite{nature}. This points to an increase in system temperature with event multiplicity. Hence, in~Figure~\ref{fig2-1}, the~x-axis becomes a replica of temperature although~the explicit functionality with event multiplicity is not known. In~view of this, 
the experimental
behavior of particle mean free path assuming a common cross section for all particle species~\cite{Kadam:2015xsa} as a function
of temperature agrees with the theoretical expectations \cite{Sarkar:2017ijd} for $q \sim 1.001$ and $\mu_{\rm B} =0$, as shown in Figures~\ref{fig1} and \ref{fig2}.
 After a threshold of $\langle dN_{\rm ch}/d\eta\rangle \sim$ (10--15), the~mean free path becomes independent of particle species, and for
 higher event multiplicities, it attains a lower asymptotic value between (1--10) fm.
 
\begin{figure}
%\begin{center}
\includegraphics[scale = 0.45]{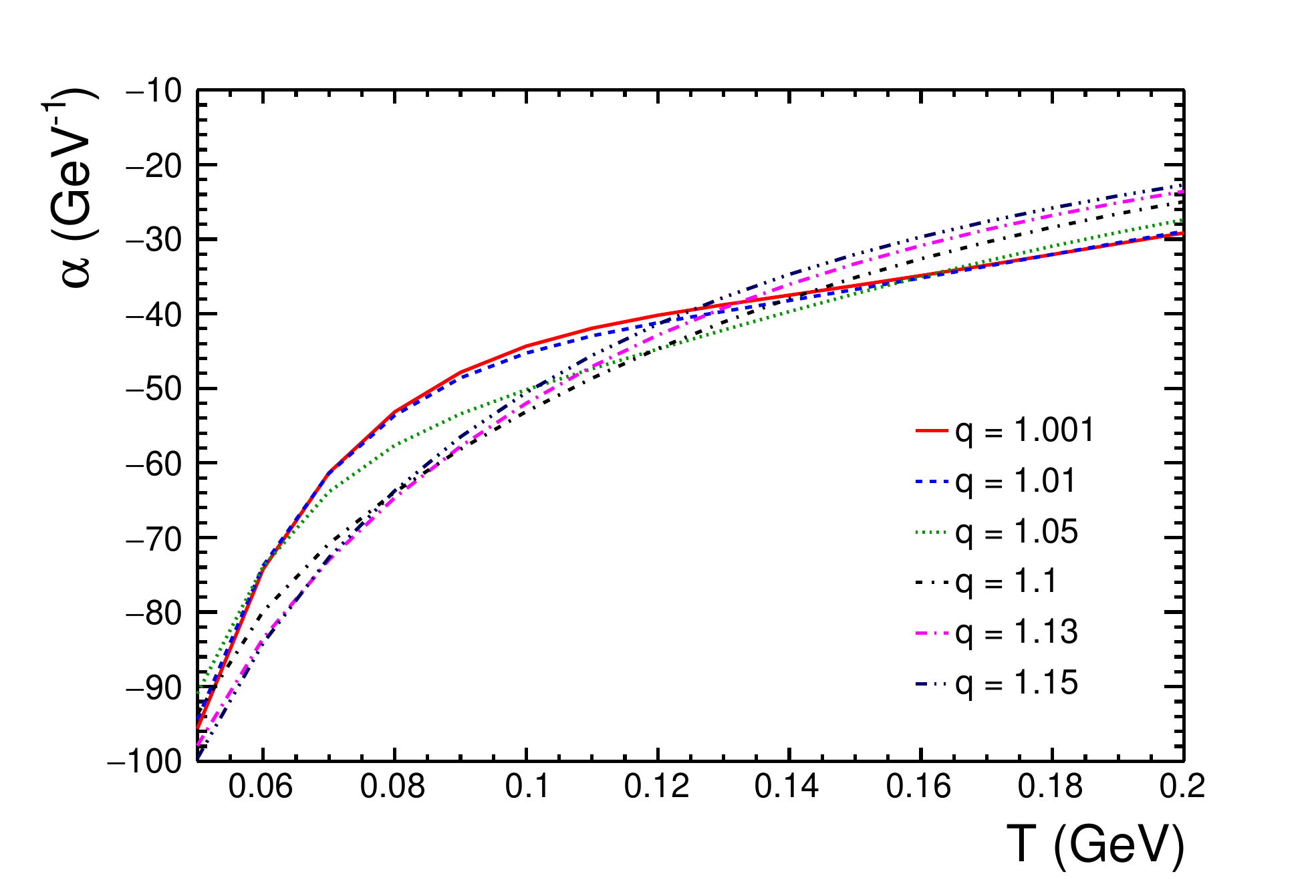}
\caption{Isobaric expansivity, $\alpha$, as a function of  temperature, 
$T$, for different values of non-extensive parameter, $q$, for hadron gas 
for baryochemical potential, $\mu_{\rm B}$ = 0.} 
\label{fig3}
%\end{center}
\end{figure}

We use Equation~(\ref{eq7}) to calculate the isobaric expansivity of a hadron gas. Figure~\ref{fig3} shows the isobaric expansivity
as a function of temperature for different values of the non-extensive parameter $q$. One observes 
that, for lower temperatures, the~system shows lower $\alpha$ values and that,~as the temperature increases slowly, $\alpha$ increases rapidly and the trend becomes almost flat for higher values of temperature, $T \sim $ 180 MeV. One can also see 
that the isobaric expansivity is highly dependent on $q$ at low temperatures. For~$q \to$ 1, which is for an equilibrated system, $\alpha$ is the lowest in the high temperature regime. For~high temperature, the~value of $\alpha$ increases as $q$ increases, i.e.,~as the system stays away from the equilibrium. The~expansivity of the system is thus found to be correlated with
the degree of non-extensivity of the system. 

\begin{figure}
%\begin{center}
\includegraphics[scale = 0.45]{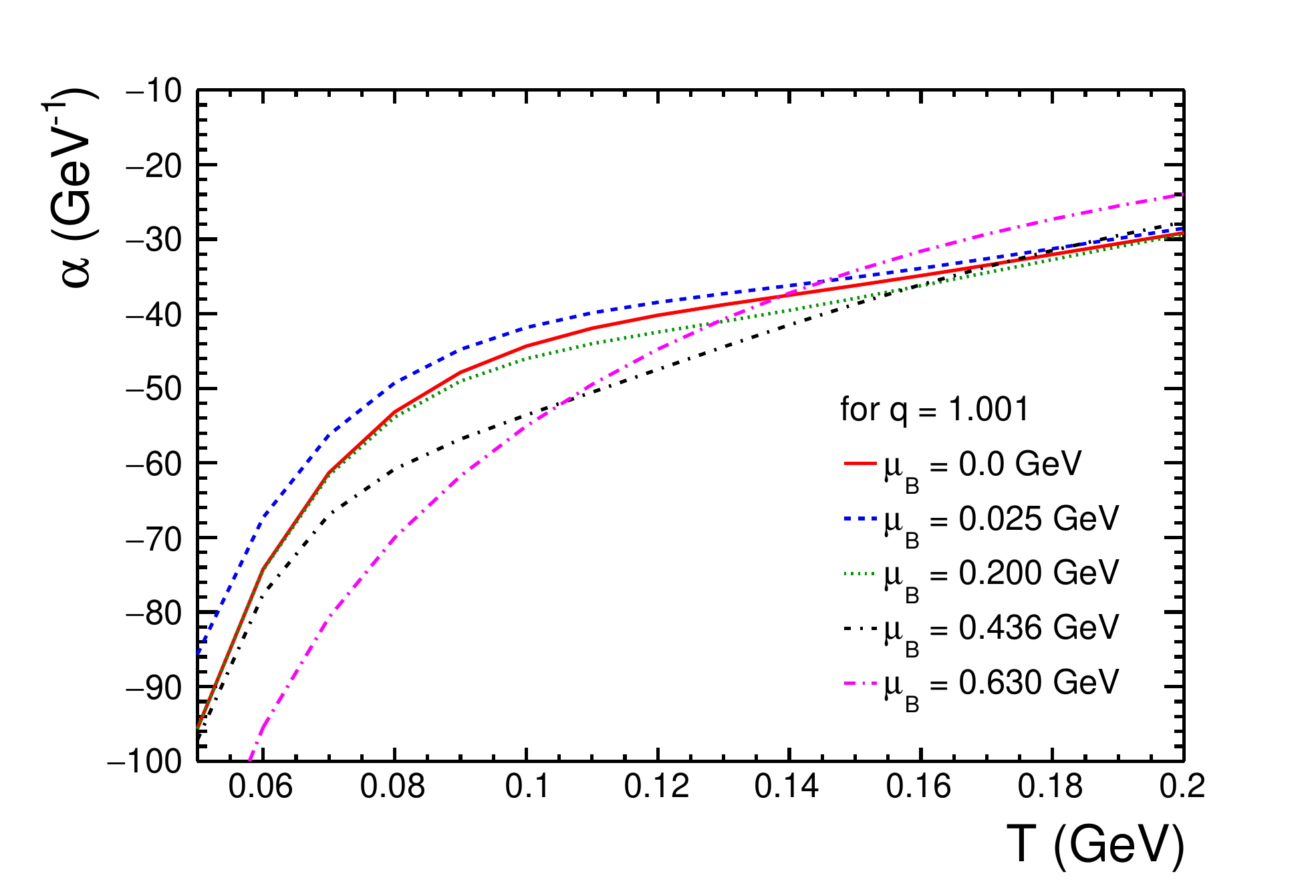}
\caption{
 Isobaric expansivity, $\alpha$, as a function of temperature, $T$, for different values of baryochemical potential, $\mu_{B}$ for hadron gas.}
\label{fig4}
%\end{center}
\end{figure}

The isobaric expansivity of hadron gas as a function of temperature for 
different values of baryochemical potential, $\mu_{\rm B}$, is shown in 
Figure~\ref{fig4}. Here, we take the non-extensive parameter $q$ = 1.001 and estimated $\alpha$ for different $\mu_{\rm B}$ values. For~$\mu_{\rm B}$ = 0 (the case at the LHC energies), we obtain a similar trend as we got for the case of  $q$ = 1.001, shown in Figure~\ref{fig3}. A~correlation between $\mu_{\rm B}$ and the expansivity of the system can be observed from Figure~\ref{fig4}. At~high temperature, the~expansivity of the hadron gas is lower for $\mu_{\rm B}$ = 0 and~it becomes higher for high $\mu_{\rm B}$ values. The~reason for negative expansivity can be due to different physical processes such as transverse vibrational modes, rigid unit modes, and phase transitions~\cite{DFisher}. However, in our case, considering an expanding fireball scenario, we know that, at higher temperatures, the system volume is smaller. As~the temperature decreases, the~volume of the system increases. Thus, we infer that, when we increase the temperature of the expanding fireball at a given point in time, its volume becomes smaller---as if it would appear to contract. This is the reason why we get a negative expansivity for a hadron~gas. 

\begin{figure}
%\begin{center}
\includegraphics[scale = 0.45]{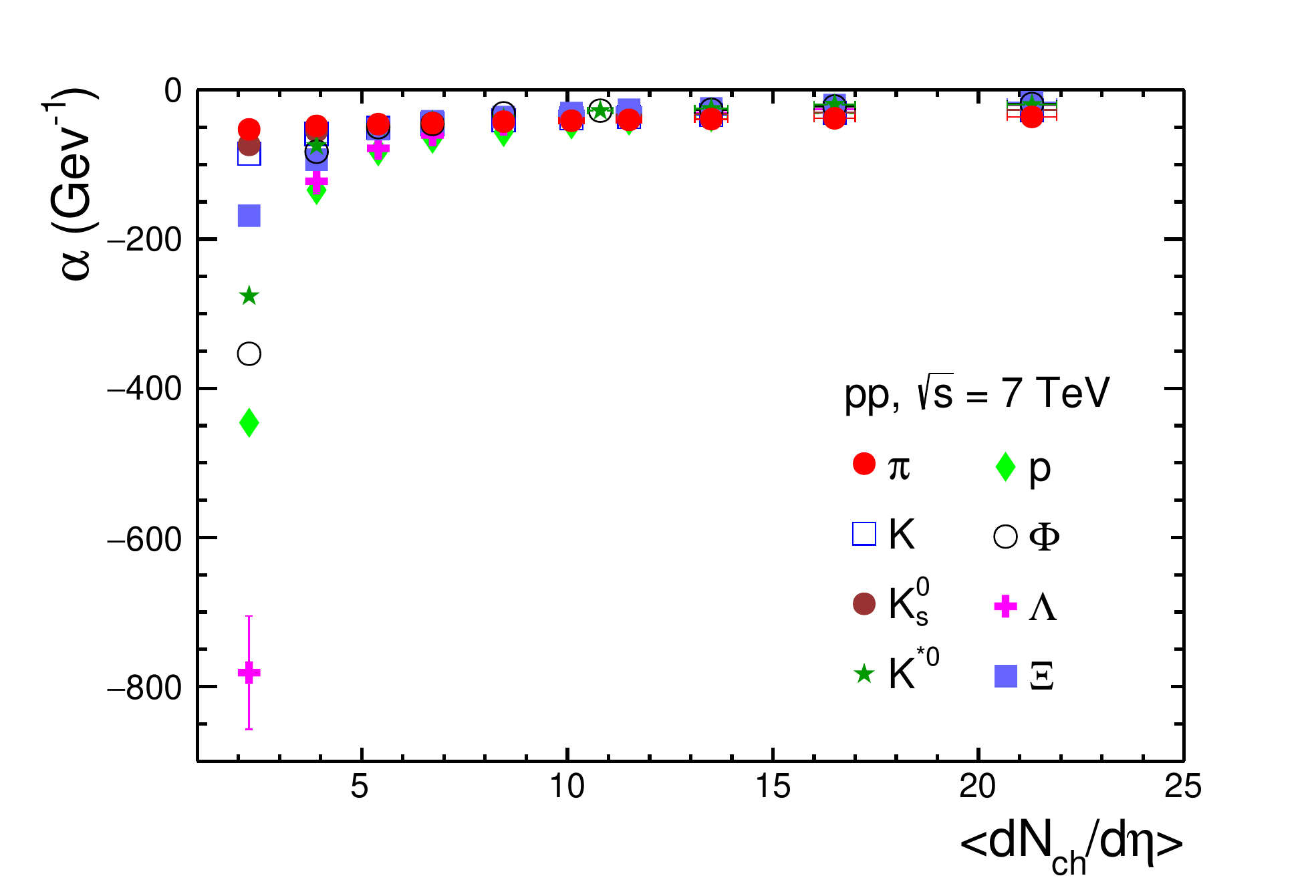}
\caption{
 Isobaric expansivity, $\alpha$, as a function of final state charged 
particle mean multiplicity density in pseudorapidity, 
 $\langle dN_{\rm ch}/d\eta\rangle$, for pp collisions at $\sqrt{s}$ = 7 
TeV for identified charged~particles. The values
of isobaric expansivity are estimated by using Equation (\ref{eq7}) by 
taking the $(T,q)$ values from the Tsallis distribution fitting to the experimental transverse momentum spectra of identified particles at the midrapidity.}
 \label{fig5}
%\end{center}
\end{figure}

Figure~\ref{fig5} shows the isobaric expansivity of the hadron gas system formed in LHC pp collisions at $\sqrt{s}$ = 7 TeV  as a function of final state-charged particle multiplicity, which is an event classifier in pp collisions. We took the temperature and $q$ values from the Tsallis fit to the $p_\perp$-spectra of the identified particles produced in LHC pp collisions at  $\sqrt{s}$ = 7 TeV~\cite{Acharya:2018orn} to estimate the isobaric expansivity of the system for different identified particles~\cite{Khuntia:2018znt}. At~the LHC energies, the~baryochemical potential of the system is almost zero, and for our studies, we use $\mu_B$ = 0. One observes an increasing trend in the expansivity of the system as the final state charged particle multiplicity increases. For~lower $\langle dN_{\rm ch}/d\eta \rangle$, the~expansivity of identified particles are different and mostly follow a mass ordering, with~massive particle having lower $\alpha$ than lighter particles. This is a consequence of the fact that, in the hadronic phase, the~number density of the lighter particles are higher than that of the massive particles---statistical production of particles in the multi-particle production process. When we increase the temperature, the~number density of all the particles increases. The~given temperature can be translated into the kinetic energy of the particles in the system. Taking the solid state and gaseous phase
 analogy, for~a denser system, the~increase in temperature makes a small increase in the volume of the system, whereas for a less dense system, the~change is very high. However, after~a certain charged particle multiplicity $\langle dN_{\rm ch}/d\eta \rangle$ = 10, we observe that all the hadronic species in the system have almost the same isobaric expansivity. This suggests a change in the dynamics of the system at $\langle dN_{\rm ch}/d\eta \rangle$ higher than 10~\cite{Sahu:2019tch}.
 
It is known that water has a negative expansivity from 0 $^{\circ}$C 
 to 4 $^{\circ}$C. However, with~the increase in temperature, it gains positive expansivity. At~30 $^{\circ}$C, water has an isobaric expansivity of about $3.515\times 10^{9} ~{\rm GeV}^{-1}$ \cite{Water}. This comparison is made to~obtain a physical realization of the system produced in high-energy~collisions.

\begin{figure}
%\begin{center}
\includegraphics[scale = 0.45]{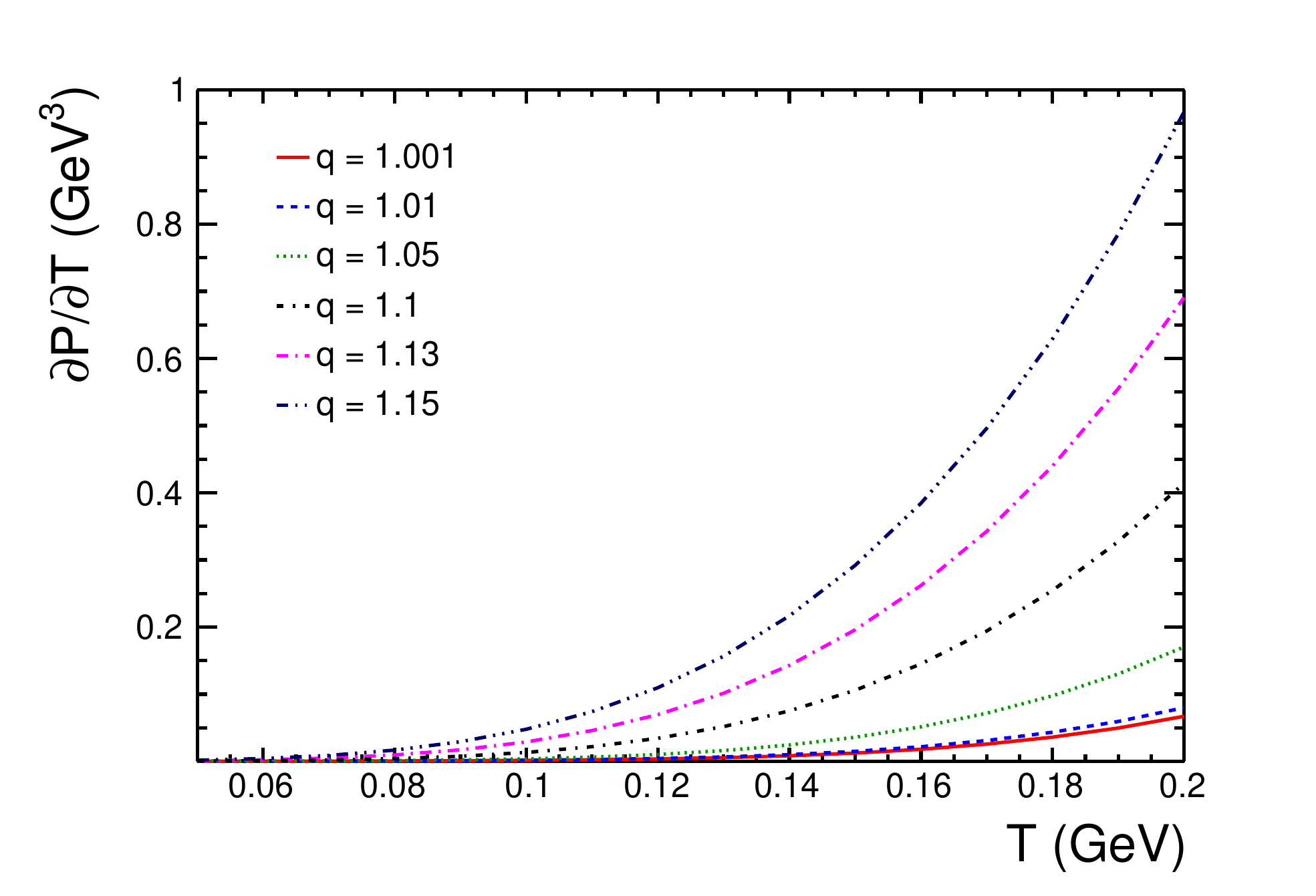}
\caption{The thermal pressure, $\frac{\partial P}{\partial T}$, of the system as a function of temperature, $T$, for hadron gas for different $q$ values for baryochemical potential, $\mu_{\rm B}$ = 0.}
\label{fig7}
%\end{center}
\end{figure}

We use Equation~(\ref{eq12}) to estimate $\frac{\partial P}{\partial T}$, the thermal pressure of the produced system. Figure~\ref{fig7} shows the variation in $\frac{\partial P}{\partial T}$ as a function of temperature for different values of the non-extensive parameter $q$. One observes that for low temperatures, the~thermal pressure is lower for the system, but~as the temperature increases, the~thermal pressure also increases. One also can see that, when the system is near equilibrium, $\frac{\partial P}{\partial T}$ gets the lowest value. For~higher $q$ values, $\frac{\partial P}{\partial T}$ increases, meaning that, as~the degree of non-extensivity of the system increases, the~thermal
pressure also increases at any given temperature. Hence, the~thermal pressure is responsible for the degree of non-equilibration
of the system. In~addition, at~lower temperatures, the~number density of the system would be smaller (as the volume of the system increases). With~a small change in temperature, the~change in pressure of the system would be smaller in comparison to the case of higher temperature regimes, where the number density of the system would be much higher, which contributes to a higher thermal~pressure.

\begin{figure}
%\begin{center}
\includegraphics[scale = 0.45]{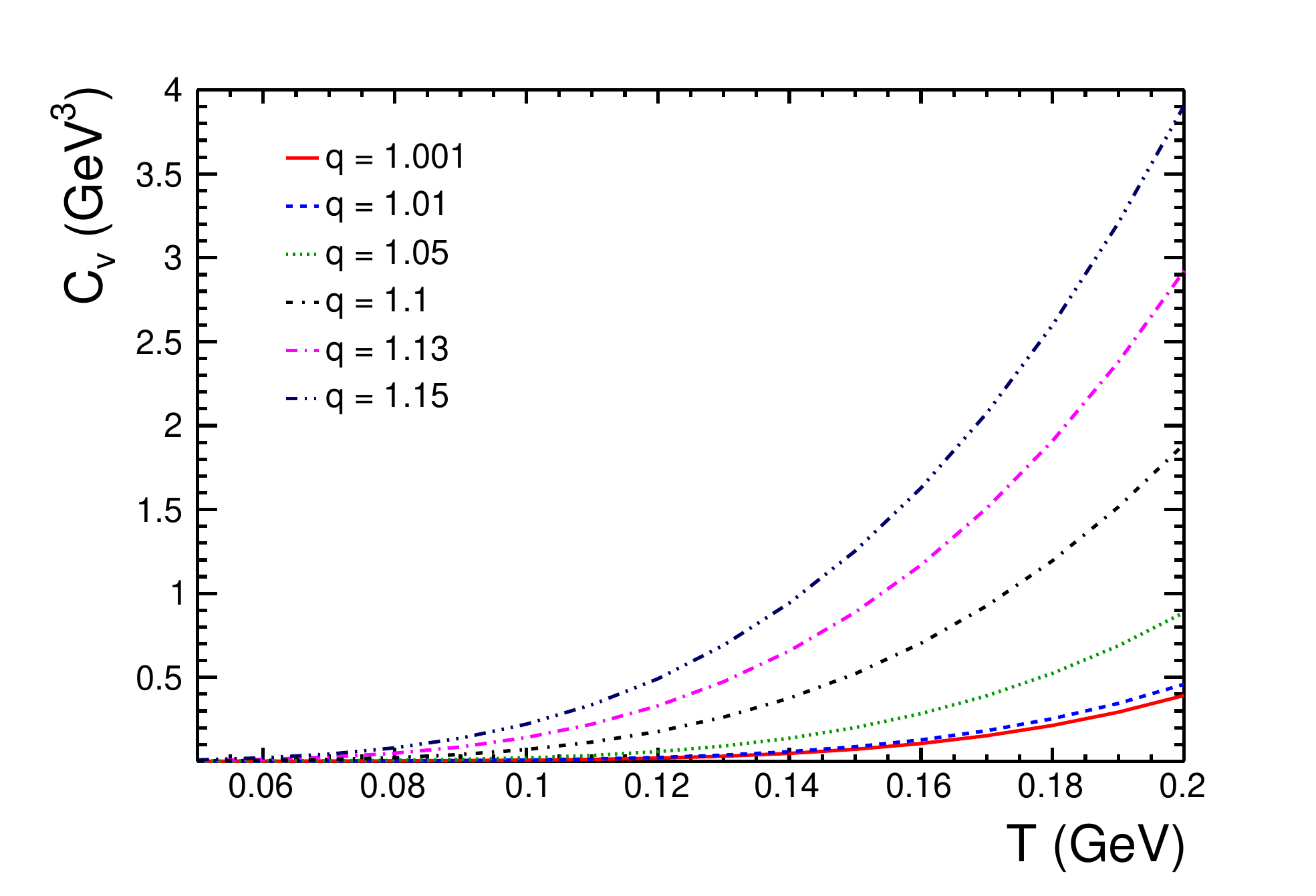}
\caption{The specific heat at constant volume, $C_{\rm V}$, as a function of temperature, $T$, for hadron gas for different $q$-values for baryochemical potential, $\mu_{\rm B}$ = 0.} 
\label{fig8}
%\end{center}
\end{figure}

\begin{figure}
%\begin{center}
\includegraphics[scale = 0.45]{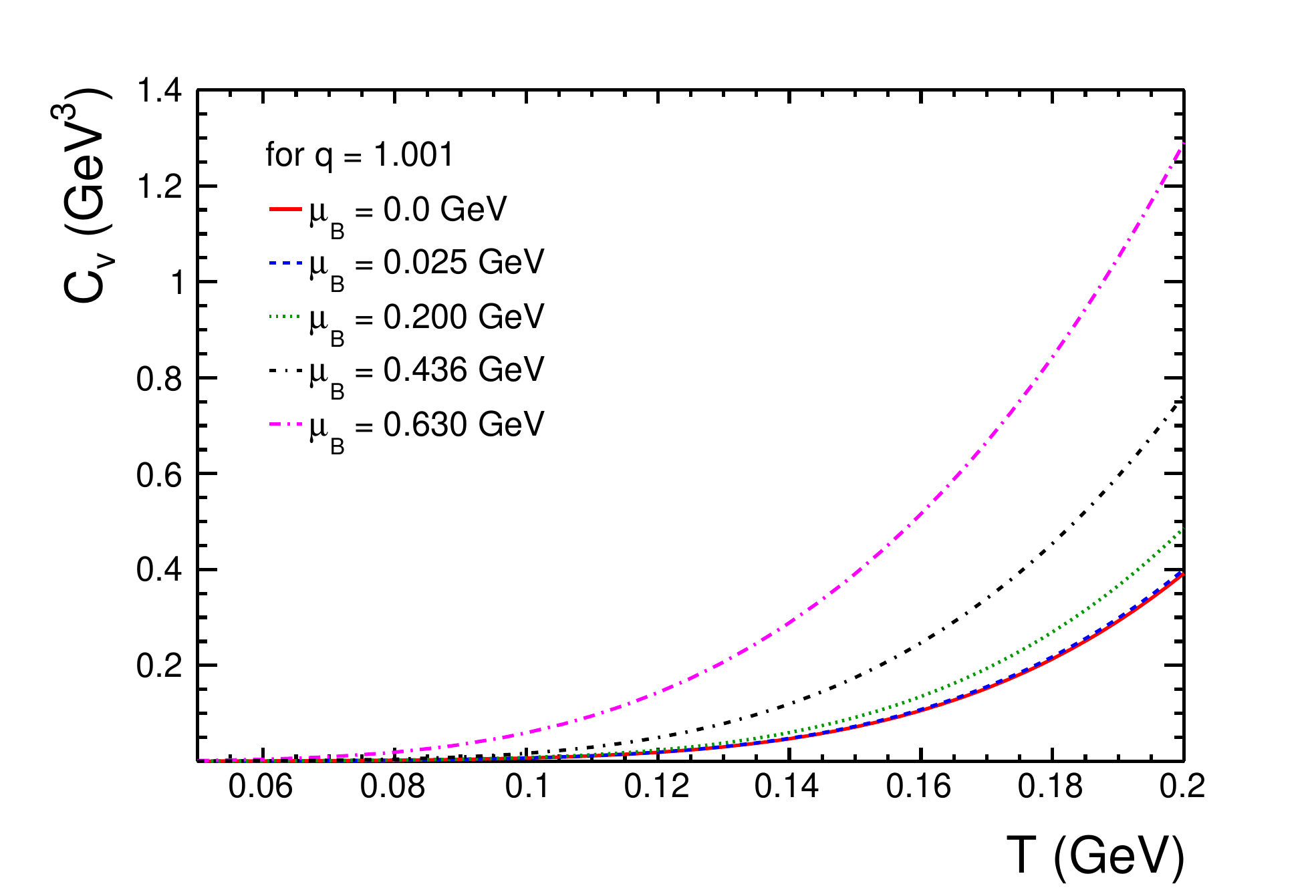}
\caption{The specific heat at constant volume, $C_{\rm V}$, as a function of temperature, $T$, for hadron gas for different baryochemical potential values, $\mu_{\rm B}$.}
\label{fig9}
%\end{center}
\end{figure}

We calculate the specific heat at constant volume for a system with hadrons as the constituents using Equation~(\ref{eq13}).
 Figure~\ref{fig8} shows the variation in $C_{\rm V}$ as a function of 
temperature. One sees that, for low temperatures, the~specific heat of the 
system is low while it increases gradually as the temperature increases. One also observes 
 that, for $q$ = 1.001, i.e.,~when the system is near equilibrium, the~specific heat of the system gets the lowest value. This is expected, as~for an equilibrated system, the degree of change in the internal energy of the system is less. For~higher $q$ values, $C_{\rm V}$  becomes higher at any given temperature. From~Figure~\ref{fig9}, one observes 
 that the baryochemical potential plays a big role in the outcome of $C_{\rm V}$. For~all $\mu_{\rm B}$ values, the~$C_{\rm V}$ of the system shows an increase.
 The higher the $\mu_{\rm B}$ values, the higher the values of specific heat at all~temperatures.
 
\unskip

\section{Summary and~Conclusions}
\label{sum}
The observations of heavy-ion-like features in high-multiplicity proton-proton 
collisions at the Large Hadron Collider energies and the corresponding explanations using different tunes in pQCD-inspired Monte Carlo model such as PYTHIA8 have warranted a deeper look into these
events. In the present study, we estimated various thermal properties of 
the matter formed in high-energy proton-proton (pp) collisions, such as 
mean free path, 
$\lambda$, isobaric expansivity, $\alpha$, thermal pressure, $\frac{\partial P}{\partial T}$, and heat capacity, $C_{\rm V}$, using a thermodynamically consistent Tsallis distribution function. The~effect of the non-extensive parameter, $q$, baryochemical potential, $\mu_{\rm B}$, and temperature, $T$, on these thermal properties are studied. Some of these observables such as mean free path and isobaric expansivity are studied as a 
function of final state charged particle multiplicity for pp collisions at the center-of-mass energy $\sqrt{s}$ = 7 TeV, and the findings are compared to the theoretical expectations. The~important results are summarized~below. 
\begin{enumerate}
\item The mean free path is found to be highly sensitive to the equilibration parameter $q$ at lower temperatures. In~the domain of high temperature, it is found to be almost independent from the $q$ parameter. The higher the baryochemical potential, the lower the mean free path of the system at a certain temperature. For~all values of baryochemical potential, the~spectra show a decreasing trend with~temperature.

\item The mean free path is then studied for identified particles with the ($T,q$) inputs from the experimental transverse momentum spectra. The  midrapidity charged particle pseudorapidity density $\langle dN_{\rm ch}/d\eta \rangle \sim$ (10--15) is found to be a threshold in the final state event multiplicity, after~which the mean free path becomes independent of particle species. For~higher event multiplicities, the~mean free path seems to attain a lower asymptotic value between (1--10) fm.

\item A negative isobaric expansivity of the hadron gas system is observed. The~isobaric expansivity is highly sensitive to the $q$ parameter. In a~higher temperature region, the~expansivity is higher for a system away from equilibrium. Similarly, for~high $\mu_{B}$ values, we observe a higher expansivity of the system in the high temperature~region.

\item For the isobaric expansivity of the identified particles, we 
observed a threshold midrapidity charged particle pseudorapidity density, $\langle dN_{\rm ch}/d\eta \rangle >$ 10, after which the isobaric expansivity almost converged to the same value. This may indicate a change in the dynamics of the~system.

\item The thermal pressure was found to be weakly dependent on the $q$ parameter at lower temperatures, but~the dependency became stronger with the increase in temperature. As~expected, at~lower temperatures, the~thermal pressure in the system is lower, which goes up, as~the temperature of the system increases. The~thermal pressure is found to be positively correlated with the degree of non-extensivity of a physical~system.

\item The specific heat at constant volume is weakly dependent on the equilibration parameter $q$ at lower temperatures and highly dependent on $q$ in the higher temperature region. As~the temperature increases, we observed an increase in the specific heat of the system. We also studied the effect of baryochemical potential on the specific heat and observed that, for a baryon-rich environment, the~specific heat becomes higher for high baryochemical~potential.

\item Although in this paper a~theoretical estimation of various important thermodynamic quantities was made as a function of final state multiplicity keeping pp collisions in mind in order to carry out a systematic study, the~physical interpretation of the thermodynamic properties of the systems corresponding to low-multiplicities must be performed with~caution.

\end{enumerate}

%%%%%%%%%%%%%%%%%%%%%%%%%%%%%%%%%%%%%%%%%%
\vspace{6pt} 

\acknowledgments{This research is carried out under DAE-BRNS Project No. 58/14/29/2019-BRNS of Government of India. D.S. acknowledges the initial discussions with Dr. Golam Sarwar and Dr. Sushanta Tripathy. Some help from Vaibhav Lohia and Archita Rani Dash on these studies are highly~appreciated.}

%%%%%%%%%%%%%%%%%%%%%%%%%%%%%%%%%%%%%%%%%%
{}


\section{References}


\begin{thebibliography}{999}

\bibitem{nature}
 ALICE Collaboration. Enhanced production of multi-strange hadrons in high-multiplicity  proton-antiproton collisions.
 \emph{Nature Phys.} \textbf{2017}, \emph{13}, 535.
  
  
\bibitem{Velicanu:2011zz} 
  Velicanu, D. (for~CMS Collaboration). Ridge correlation structure in 
high-multiplicity pp collisions with CMS. \emph{J.\ Phys.\ G} \textbf{2011}, {\em 38}, 124051.
  
   
\bibitem{Pythia8} Sjostrand, T.; Mrenna, S.;  Skands, P. Z. PYTHIA 6.4 
physics and manual
\emph{JHEP} \textbf{2006}, {\em 05}, 026.

 %\cite{Ortiz:2013yxa}
\bibitem{Ortiz:2013yxa}
Ortiz Velasquez, A.; Christiansen, P.; Flores, E.C.; Cervantes, I.M.; 
Pai\'c, 
G.
Color reconnection and flowlike patterns in pp collisions.
\emph{Phys. Rev. Lett.} \textbf{2013}, \emph{111}, 042001.

\bibitem{Nayak:2018xip}
Nayak, R.; Pal, S.; Dash, S.
Effect of rope hadronization on strangeness enhancement in p$-$p collisions at LHC energies.
\emph{Phys. Rev. D} \textbf{2019}, \emph{100}, 074023.

\bibitem{Werner:2013tya}
Werner, K.; Guiot, B.; Karpenko, I.; Pierog, T.
Analysing radial flow features in p-Pb and p-p collisions at several TeV by studying identified particle production in EPOS3.
\emph{Phys. Rev. C} \textbf{2014}, \emph{89}, 064903.

\bibitem{Werner:2010aa}
Werner, K.; Karpenko, I.; Pierog, T.; Bleicher, M.; Mikhailov, K.
Event-by-event simulation of the three-dimensional hydrodynamic evolution 
from flux tube initial conditions in ultrarelativistic heavy ion 
collisions.
\emph{Phys. Rev. C} \textbf{2010}, \emph{82}, 044904.

\bibitem{Drescher:2000ha}
Drescher, H.; Hladik, M.; Ostapchenko, S.; Pierog, T.; Werner, K.
Parton based Gribov-Regge theory.
\emph{Phys. Rept.} \textbf{2001}, \emph{350}, 93.


\bibitem{Acharya:2019mzb}
Acharya, S.; AdamovÃ¡, D.; Adhya, S.P.; Adler, A.; Adolfsson, J.; Aggarwal, M.M.;  Rinella, G.A.; Agnello, M.; Agrawal, N.; Ahammed, Z.; et al.  
Charged-particle production as a function of multiplicity and transverse spherocity in pp collisions at $\sqrt{s} =5.02$ and 13 TeV.
\emph{Eur. Phys. J. C} \textbf{2019}, \emph{79}, 857.

  
\bibitem{Khatun:2019dml}
Khatun, A.; Thakur, D.; Deb, S.; Sahoo, R.
$J/\psi$ Production dynamics: Event shape, multiplicity and rapidity 
dependence in proton+proton collisions at LHC energies using PYTHIA8.
\emph{J. Phys. G} \textbf{2020}, \emph{47}, 055110.

%\cite{Cuautle:2015kra}
\bibitem{Cuautle:2015kra}
Ortiz, A.; Paic, G.; Cuautle, E.
Mid-rapidity charged hadron transverse spherocity in pp collisions simulated with Pythia.
\emph{Nucl. Phys. A} \textbf{2015}, \emph{941}, 78.

%\cite{Rath:2019izg}
\bibitem{Rath:2019izg}
Rath, R.; Khuntia, A.; Tripathy, S.; Sahoo, R.
A Baseline Study of the Event-shape and multiplicity dependence of 
chemical freeze-out parameters in proton-proton collisions at $\sqrt{s}$ = 
13 TeV using PYTHIA8.
\emph{Physics} \textbf{2020}, \emph{2}, 679.

 %\cite{Tripathy:2019blo}
\bibitem{Tripathy:2019blo}
Tripathy, S.; Bisht, A.; Sahoo, R.; Khuntia, A.; Malavika,~P.S.
Event shape and multiplicity dependence of freeze-out scenario and system 
thermodynamics in proton+proton collisions at $\sqrt{s}$ = 13 TeV Using 
PYTHIA8. \emph{ Adv. High Energy Phys.} \textbf{2021},  \emph{2021}, 8822524
 
  %\cite{Khuntia:2018qox}
\bibitem{Khuntia:2018qox}
Khuntia, A.; Tripathy, S.; Bisht, A.; Sahoo, R. Event shape engineering 
and multiplicity dependent study of identified particle production in 
proton+proton collisions at $\sqrt{s}$ = 13 TeV using PYTHIA.
\emph{J. Phys. G} \textbf{2021}, \emph{48}, 035102.
  
  %\cite{Deb:2019yjo,Azmi:2019irb}
\bibitem{Deb:2019yjo}
Deb, S.; Sarwar, G.; Sahoo, R.; Alam, J.-e. 
Study of QCD dynamics using small systems. \emph{arXiv} \textbf{2019},
arXiv:1909.02837.

%\cite{Azmi:2019irb}
\bibitem{Azmi:2019irb}
Azmi, M.; Bhattacharyya, T.; Cleymans, J.; Paradza, M.
Energy density at kinetic freeze-out in Pb-Pb collisions at the LHC using the Tsallis distribution.
\emph{J. Phys. G} \textbf{2020}, \emph{47}, 045001.
  
 %Add here  
  \bibitem{Bhattacharyya:2017hdc}
Bhattacharyya, T.; Cleymans, J.; Marques, L.; Mogliacci, S.; Paradza, M.~W.
On the precise determination of the Tsallis parameters in proton\textendash{}proton collisions at LHC energies.
\emph{J. Phys. G} \textbf{2018}, {\em 45}, 055001. 

\bibitem{Khuntia:2017ite}
Khuntia, A.; Tripathy, S.; Sahoo, R.; Cleymans, J.
Multiplicity dependence of non-extensive parameters for strange and 
multi-strange particles in proton-proton collisions at $\sqrt{s}= 7$ TeV 
at the LHC.
\emph{Eur. Phys. J. A} \textbf{2017}, {\em 53}, 103.
   
\bibitem{KHuang}
  Huang, K. \emph{Statistical  Mechanics}; John Wiley: New York, 
NY, USA, 1987.

\bibitem{Kharzeev:2017qzs}
Kharzeev, D. E. and Levin, E. M. Deep inelastic scattering as a probe of entanglement. \emph{Phys. Rev. D} \textbf{2017}, {\em 95}, 114008.
  
\bibitem{Abelev:2006cs} 
  Abelev, B.I.; Adams, J.; Aggarwal, M.M.; Ahammed, Z.; Amonett, J.; Anderson, B.D.; Anderson, M.; Arkhipkin, D.; \mbox{Averichev, G.S.;} Bai, Y.; et al. 
Strange particle production in p+p collisions at $\sqrt{s}$ = 200 GeV.
  \emph{Phys.\ Rev.\ C} \textbf{2007}, {\em 75}, 064901.


\bibitem{Adare:2011vy} 
  Adare, A.; Afanasiev, S.; Aidala, C.; Ajitan, ; N.N.; Akiba, Y.; Al-Bataineh, H.; Alexander, J.; Aoki, K.; Aphecetche, L.; \mbox{Armendariz, R.;~et~al.} 
Identified charged hadron production in $p+p$ collisions at $\sqrt{s}=200$ and 62.4 GeV.
  \emph{Phys.\ Rev.\ C} \textbf{2011}, {\em 83}, 064903.
  
  
\bibitem{Aamodt:2011zj} 
  Aamodt, K.; Abel, N.; Abeysekara, U.; Quintana, A.A.; Abramyan, A.; Adamova, D.; Aggarwal, M.M.; Rinella, G.A.; Agocs, A.G.; Salazar, S.A.;~et~al. 
Production of pions, kaons and protons in pp collisions at $\sqrt{s}= 900$ GeV with ALICE at the LHC.
  \emph{Eur.\ Phys.\ J.\ C} \textbf{2011}, {\em 71}, 1655.
  
  
\bibitem{Abelev:2012cn} 
  Abelev, B.; Quintana, A.A.; AdamovÃ¡, D.; Adare, A.M.; Aggarwal, M.M.; Rinella, G.A.; Agocs, A.G.; Agostinelli, A.; Salazar, S.A.; Ahammed, Z.; ~et~al. 
Neutral pion and $\eta$ meson production in proton-proton collisions at $\sqrt{s}=0.9$ TeV and $\sqrt{s}=7$ TeV.
  \emph{Phys.\ Lett.\ B} \textbf{2012}, {\em 717}, 162.
  
  
\bibitem{Abelev:2012jp} 
 ALICE Collaboration.
Multi-strange baryon production in $pp$ collisions at $\sqrt{s} = 7$ TeV 
with ALICE.
  P\emph{hys.\ Lett.\ B} \textbf{2012}, {\em 712}, 309.
  
  
\bibitem{Chatrchyan:2012qb} 
  Chatrchyan, S.; Khachatryan, V.; Sirunyan, A.M.; Tumasyan, A.; Adam, W.; 
Aguilo, E.; Bergauer, T.; Dragicevic, M.; ErÃ¶, J.; \mbox{Fabjan, 
C.;~et~al.}  Study of the inclusive production of charged pions, kaons, 
and 
protons in $pp$ collisions at $\sqrt{s}=0.9$, 2.76, and 7 TeV.
  \emph{Eur.\ Phys.\ J.\ C} \textbf{2012}, {\em 72}, 2164.
  
  \bibitem{Bhattacharyya:2015hya}
Bhattacharyya, T.; Cleymans, J.; Khuntia, A.; Pareek, P.; Sahoo, R. 
Radial flow in non-extensive thermodynamics and study of particle spectra 
at LHC in the limit of small $(q-1)$. \emph{Eur. Phys. J. A.} 
\textbf{2016}, {\em 52}, 30. 
  
\bibitem{Pirner:2011ab}
Pirner, H.J.; Reygers, K.
Light-cone QCD plasma.
\emph{Phys. Rev. D} \textbf{2012}, \emph{86}, 034005.

 
%\bibitem{Tsallis:52}
%  Tsallis, C. Possible generalization of Boltzmann-Gibbs statistics.
% \emph{J. Stat. Phys.} \textbf{1988}, \emph{52}, 479.


%\cite{Cleymans:2011in}
\bibitem{Cleymans:2011in}
Cleymans, J.; Worku, D.
The Tsallis distribution in proton-proton collisions at $\sqrt{s}$ = 0.9 
TeV at the LHC.
\emph{J. Phys. G} \textbf{2012}, \emph{39}, 025006.

 
\bibitem{Cleymans:2012ya} 
  Cleymans, J.; Worku, D.
Relativistic thermodynamics: transverse momentum distributions in 
high-energy physics.
  \emph{Eur.\ Phys.\ J.\ A} \textbf{2012}, {\em 48}, 160.
  
\bibitem{Deppman} 
  Deppman, A. Self-consistency in non-extensive thermodynamics of highly excited hadronic states.
 \emph{Physica A} \textbf{2012}, {\em 391}, 6380.
   
\bibitem{Deppman1} 
  Deppman, A. Properties of hadronic systems according to the nonextensive self-consistent thermodynamics.
   \emph{J. Phys. G Nucl. Part. Phys.} \textbf{2014}, {\em 41}, 055108.
  
  %\cite{Biro:2020kve}
\bibitem{Biro:2020kve}
B\'ir\'o, G.; Barnaf\"{o}ldi, G.G.; Bir\'o, T.S.
Tsallis-thermometer: A QGP indicator for large and small collisional systems. \emph{arXiv} \textbf{2020}, 
arXiv:2003.03278.

\bibitem{Deppman:2017fkq}
Deppman, A.; Megias, E.; Menezes, D.P.; Frederico, T.
Fractal structure and non extensive statistics.
\emph{Entropy} \textbf{2018}, \emph{20}, 633.
  
 
\bibitem{Kadam:2015xsa} 
  Kadam, G.P.; Mishra, H.
Dissipative properties of hot and dense hadronic matter in an excluded-volume hadron resonance gas model.
  \emph{Phys.\ Rev.\ C} \textbf{2015}, {\em 92}, 035203.
   
 
\bibitem{Bugaev} Bugaev, K.A.; Oliinychenko, D.R.; Sorin, A.S.; Zinovjev, 
G.M. Simple solution to the strangeness horn description puzzle. 
 \emph{Eur.\ Phys.\ J. A} \textbf{2013}, {\em 49}, 30.

\bibitem{PDG} Zyla, P.A. {\it et al.} (Particle Data Group), Review of Particle Physics. \emph{Prog. Theor. Exp. Phys.} {\bf 2020}, {\it 2020}, 083C01.

\bibitem{Cleymans:2020ojr}
Cleymans, J.; Paradza, M.W.
Tsallis statistics in high energy physics: Chemical and thermal 
freeze-outs.
\emph{Physics} \textbf{2020}, \emph{2}, 654.

  
 
\bibitem{Kraus:2007hf}
Kraus, I.; Cleymans, J.; Oeschler, H.; Redlich, K.; Wheaton, S.
Chemical equilibrium in collisions of small systems.
\emph{Phys. Rev. C} \textbf{2007}, \emph{76}, 064903.

\bibitem{Tawfik:2016sqd} Tawfik, N.A.; Abou-Salem, L.I.; Shalaby, A.G.; Hanafy, M.; Sorin, A.; Rogachevsky, O.; Scheinast, W. Particle production and chemical freezeout from the hybrid UrQMD approach at NICA energies.
  \emph{Eur.\ Phys.\ J.\ A} \textbf{2016}, {\em 52}, 324.

  
\bibitem{BraunMunzinger:2001ip} Braun-Munzinger, P.; Magestro, D.; Redlich, K.; Stachel, J. Hadron production in Au-Au collisions at RHIC.
  \emph{Phys.\ Lett.\ B} \textbf{2001}, {\em 518}, 41.

 
  
\bibitem{Cleymans:2005xv} Cleymans, J.; Oeschler, H.; Redlich, K.; Wheaton, S. 
Comparison of chemical freeze-out criteria in heavy-ion collisions.
  \emph{Phys.\ Rev.\ C} \textbf{2006}, {\em 73}, 034905.


\bibitem{Khuntia:2018non} Khuntia, A.; Tiwari, S.K.; Sharma, P.; Sahoo, R.; Nayak, T.K.
Effect of Hagedorn states on isothermal compressibility of hadronic matter 
formed in heavy-ion collisions: From NICA to LHC energies.
  \emph{Phys.\ Rev.\ C} \textbf{2019}, {\em 100}, 014910.

\bibitem{Khuntia:2018znt} 
  Khuntia, A.; Sharma, H.; Tiwari, S.K.; Sahoo, R.; Cleymans, J.
Radial flow and differential freeze-out in proton-proton collisions at $\sqrt{s} = 7$ TeV at the LHC.
  \emph{Eur.\ Phys.\ J.\ A} \textbf{2019}, {\em 55}, 3.  
   
   \bibitem{Sarkar:2017ijd}
Sarkar, N.; Ghosh, P. Thermalization in a small hadron gas system and high-multiplicity $pp$ events. 
\emph{Phys. Rev. C} \textbf{2017}, {\em 96}, 04490.
   
\bibitem{DFisher}
 Fisher, D.J. \emph{Negative Thermal Expansion Materials};

 Materials Research Forum LLC: Millersville, PA, USA, 2018. 
 https://doi.org/10.21741/9781945291494
  
    %\cite{Acharya:2018orn}
\bibitem{Acharya:2018orn} 
  ALICE Collaboration.
  Multiplicity dependence of light-flavor hadron production in pp collisions at $\sqrt{s}$ = 7 TeV.
  \emph{Phys.\ Rev.\ C} \textbf{2019}, {\em 99}, 024906.
  
\bibitem{Sahu:2019tch} 
  Sahu, D.; Tripathy, S.; Pradhan, G.S.; Sahoo, R.
Role of event multiplicity on hadronic phase lifetime and QCD phase boundary in ultrarelativistic collisions at energies available at the BNL Relativistic Heavy Ion Collider and CERN Large Hadron Collider.
  \emph{Phys.\ Rev.\ C} \textbf{2020}, {\em 101}, 014902.
    
    
\bibitem{Water}
  Kell, G.S. Density, thermal expansivity, and compressibility of liquid water from 0. deg. to 150. deg.. Correlations and tables for atmospheric pressure and saturation reviewed and expressed on 1968 temperature scale.
 \emph{J. Chem. Eng. Data} \textbf{1975}, {\em 20}, 97.
   
 \end{thebibliography}
\end{document}